\documentclass[twocolumn,showpacs,preprintnumbers,amsmath,amssymb,
floatfix,nofootinbib]{revtex4}

\usepackage{amsmath,amsfonts,bbm,euscript,hyperref}
\usepackage{graphicx}
\usepackage{dcolumn}
\usepackage{bm}
\usepackage{epsfig}
\epsfclipon

\newcommand{\nco}{\newcommand}
\nco{\beq}{\begin{equation}} \nco{\eeq}{\end{equation}}
\nco{\beqa}{\begin{eqnarray}} \nco{\eeqa}{\end{eqnarray}}
\def\be{\begin{equation}}
\def\ee{\end{equation}}    
\def\baray{\begin{eqnarray}}
\def\earay{\end{eqnarray}}

\nco{\lra}{\leftrightarrow}

\nco{\sss}{\scriptscriptstyle} \nco{\dphi}{\varphi}
\nco{\lsim}{\mbox{\raisebox{-.6ex}{~$\stackrel{<}{\sim}$~}}}
\nco{\gsim}{\mbox{\raisebox{-.6ex}{~$\stackrel{>}{\sim}$~}}}

\def\IK{\relax{\rm I\kern-.20em K}}
\def\IM{\relax{\rm I\kern-.20em M}}

\def\subo{{(1)}}
\def\subt{{(2)}}

\def\subz{{(0)}}

\def\lsim{\mbox{\raisebox{-.6ex}{~$\stackrel{<}{\sim}$~}}}
\def\gsim{\mbox{\raisebox{-.6ex}{~$\stackrel{>}{\sim}$~}}}
\def\sss{\scriptscriptstyle}

\def\done{\delta^{(1)}}
\def\dtwo{\delta^{(2)}}
\def\sH{\mathcal{H}}
\def\Mpl{M_p}
\def\Lap{\partial^{k}\partial_{k}}
\def\Linv{\triangle^{-1}}

\def\Grad{\vec{\nabla}}

\begin{document}


\title{Nongaussianity from Tachyonic Preheating in Hybrid Inflation}

\author{Neil Barnaby, James M.\ Cline}

\affiliation{%
\centerline{Physics Department, McGill University,
3600 University Street, Montr\'eal, Qu\'ebec, Canada H3A 2T8}
e-mail: barnaby@hep.physics.mcgill.ca, jcline@physics.mcgill.ca }

\date{November, 2006}

\begin{abstract} 

In a previous work we showed that large nongaussianities and
nonscale-invariant distortions in the CMB power spectrum can be
generated in hybrid inflation models, due to the contributions of the
tachyon (waterfall) field to the second order curvature
perturbation.  Here we clarify, correct, and extend those results. 
We show that large nongaussianity occurs only when the tachyon
remains light throughout inflation, whereas $n=4$ contamination to
the spectrum is the dominant effect when the tachyon is heavy.  We
find  constraints on the parameters of warped-throat
brane-antibrane inflation from nongaussianity.  For F-term and D-term
inflation models from supergravity, we obtain nontrivial constraints
from the spectral distortion effect.  We also establish that our 
analysis applies to complex tachyon fields.  

\end{abstract}

\maketitle

\section{Introduction} In the simplest models of inflation, the
primordial density perturbations have a negligible degree of
nongaussianity.  The parameter $f_{NL}$ which characterizes
nongaussianity is  of the order of $|n-1|\ll 1$ (where $n$ is  the
spectral index) in conventional inflation models
\cite{BMR}-\cite{SeeryLidsey},  whereas the current experimental
limit is $|f_{NL}|\lsim 100$ \cite{WMAP3}; one can additionally
characterize the nongaussianity using the trispectrum, which is also
small in conventional models \cite{trispectrum}.    Nevertheless,
there have been  intense theoretical efforts to find models which
predict observably large levels \cite{otherNG} (see \cite{NGreview}
for a review).   It has been difficult to find examples which give
large $f_{NL}$.  In single field inflation models a small inflaton
sound speed is necessary to achieve large nongaussianity
\cite{large_nongauss}, as in the models of \cite{DBI,leblond}, unless the
inflaton potential has a sharp feature \cite{feature}.  

The simplest multi-field  models do not seem to give large 
nongaussianity \cite{multi_field}, though it is not clear if this is
true also of more  complicated models.  Thus it is quite significant
that one of the most prevalent classes of models, hybrid inflation
\cite{hybrid}, is able to yield large nongaussianity for certain
ranges of parameters \cite{BC}.  The effect is due to the growth of the
waterfall field---tachyonic  preheating---which contributes to the
curvature perturbation, and hence the temperature anisotropy, only
starting at second order in cosmological perturbation theory.
We show that, depending on the values of certain model parameters,
two interesting effects are possible: $n=4$ distortion
of the spectrum or large nongaussianity.  The same effect was observed 
in \cite{preheatNG0}, though tachyonic preheating after hybrid inflation
was not considered in that paper.\footnote{Reference \cite{preheatNG0}
studied the behaviour of metric perturbations during preheating
in the model $V = \lambda \phi^4 / 4 + g^2\phi^2\chi^2/2 + \lambda' \chi^4/4$
and found that an $n=4$ contamination of the spectrum is generated
when the $\chi$ field is heavy throughout inflation while large nongaussianity
is possible when $\chi$ field is light.  In this paper we consider a different
model, tachyonic preheating, finding results which are qualitatively
similar.}
Nongaussianity from preheating has also been studied
in \cite{preheatNG1}-\cite{preheatNG6}.   The calculations are
complicated, and required numerical integrations over time and
wavenumbers; hence the results are not immediately intuitive.   One
of our goals in the present paper is to give a better understanding
of this novel effect, and to present some new results concerning the
application of these  results to popular models of hybrid inflation
including brane inflation \cite{KKLMMT} and P-term inflation
\cite{Pterm}, which is a synthesis of supergravity inflationary
models interpolating between F-term and D-term inflation.

We begin by reviewing the results of \cite{BC} in section \ref{II}.
In section \ref{III} we apply these results by establishing
constraints on the parameters of hybrid inflation, coming either from
the production of large nongaussianity, or from nonscale-invariant
contributions to the spectrum (as opposed to bispectrum).  These
results extend and correct our previous limits \cite{BC}.  We then
adapt them to the cases of brane-antibrane inflation in section
\ref{IV} and P-term inflation in section \ref{V}.  We further extend
our analysis to the more realistic case of a complex tachyon field
in section \ref{VI}, showing that
the extra components of the tachyon add in a simple way and amplify
the real-field results by factors of order unity.  Conclusions are
given in section \ref{VII}.  Appendix A gives details about the
matching between early- and late-time WKB solutions of the tachyon
fluctuation mode functions, while appendix B gives details about
the source term of the curvature perturbation for complex tachyons.

\section{Review of previous results}
\label{II}
\subsection{Hybrid inflation}
The hybrid inflation model which we study is defined by the 
potential
\begin{equation}
\label{pot}
  V(\varphi,\sigma) = 
\frac{\lambda}{4} \left( \sigma^2 - v^2 \right)^2 
+ \frac{m_\varphi^2}{2}\varphi^2 + \frac{g^2}{2} \varphi^2 \sigma^2
\end{equation}
where $\varphi$ is the inflaton and $\sigma$ is the tachyonic field.
Its mass depends on $\varphi$ as 
$m_\sigma^2 = -\lambda v^2 + g^2 \varphi^2$, which changes sign
when $\varphi$ reaches the critical value 
$\varphi_c = (\sqrt{\lambda}/{g}) v$. At this time, fluctuations
in the tachyon field start to grow exponentially.
This phase of exponential growth is called tachyonic preheating 
\cite{tachyonic1}-\cite{GarciaBellido};  see also \cite{preheating}
for a discussion of the general theory of preheating and
\cite{resonance} for a different type of tachyonic preheating.  

During the early stages of preheating, before the fluctuations have
become nonperturbatively large and before the backreaction has set
in, the expansion of the universe will still be  approximately de
Sitter.  Once the tachyon fluctuations become sufficiently large
their backreaction modifies the expansion of the universe and brings
inflation to an end.  This happens at a time $N_\star \equiv H
t_\star$ when the fluctuations in $\sigma$ grow to a certain value,
\begin{equation}
\label{end_of_inflation}
\left\langle \delta\sigma^2(N_*)\right\rangle 
=   \left. \int \frac{d^3k}{(2\pi)^3} |\xi_k|^2 \right|_{N = N_\star}
 = \frac{v^2}{4}
\end{equation}
where $\xi_k$ is the mode function for the fluctuations (discussed
below).  This happens at some time after the onset of the
instability. For a wide range of parameters (including the values
originally considered in \cite{hybrid}) one has $N_\star \ll 1$ so
that the symmetry breaking completes on a time scale short compared
to the Hubble time.  This is the usual {\it waterfall condition} regime of
hybrid inflation.  In the present work we will consider both the
possibilities that $N_\star \ll 1$ and also  $N_\star \gsim 1$.

We
find it convenient to measure time in terms of number of e-foldings,
taking $N=0$ to coincide with the onset of the instability,
when $m_\sigma^2=0$, and $N_*$
to be the end of inflation, defined by (\ref{end_of_inflation}).
Horizon crossing occurs at some $N_i<0$, so the number of e-foldings
since horizon crossing is $N_e = N_* - N_i$.  We determine $N_e$
using the standard relation
\begin{equation}
\label{Ne2}
  N_e = 62 - \ln\left(\frac{10^{16}\ \mathrm{GeV}}{V^{1/4}}\right) 
  - \frac{1}{3}\ln\left(\frac{V^{1/4}}{\rho_{\mathrm{r.h.}}}\right)
\end{equation}
with the energy density at reheating ($\rho_{\mathrm{r.h.}}\sim
T_{\mathrm{r.h.}}^4$) assumed to be limited by the gravitino bound
$T_{\mathrm{r.h.}}\lsim 10^{10}$ GeV, though we have checked that this
assumption has little effect on our results.  Given $N_*$ and $N_e$, 
$N_i$ is determined by $N_i = N_*-N_e$.

\subsection{Second order curvature perturbation}

We work up to second order in perturbation theory, employing the longitudinal gauge
throughout.  The expanded metric and Einstein equations can be found in \cite{BC}.
The matter content of the theory is expanded in perturbation theory as
\begin{eqnarray*}
  \varphi(\tau,\vec{x}) &=& \varphi_0(\tau) + \done \varphi(\tau,\vec{x}) 
                            + \frac{1}{2}\dtwo \varphi(\tau,\vec{x}) \\
  \sigma(\tau,\vec{x}) &=& \done \sigma(\tau,\vec{x}) 
                            + \frac{1}{2}\dtwo \sigma(\tau,\vec{x})
\end{eqnarray*}
As discussed in \cite{BC}, we are justified in dropping the homogeneous backgound
of the tachyon field $\langle \sigma(\tau,\vec{x}) \rangle \equiv \sigma_0(\tau) = 0$.
Conformal time, $\tau$, is related to cosmic time as $dt = ad\tau$.  We denote derivaties
with respect to conformal time as $f' = \partial_\tau f$ and with respect to cosmic time
as $\dot{f} = \partial_t f$.

Similarly the gauge invariant curvature perturbation, $\zeta$, is expanded in perturbation
theory as
\[
  \zeta = \zeta^\subo + \frac{1}{2}\zeta^\subt
\]
Because $\sigma_0 = 0$ the first order contribution $\zeta^\subo$ is identical to the
standard result from single field models.  We split the second order curvature perturbation
into a component which is due to the inflaton field and a component which is due to the
tachyon field as
\[
  \zeta^\subt = \zeta^\subt_\varphi + \zeta^\subt_\sigma
\]
The second order inflaton curvature perturbation, $\zeta^\subt_\varphi$, coincides with the 
$\zeta^\subt$ in single field models.  This contribution has been previously computed and
is known to be small and conserved on large scales \cite{BMR}-\cite{SeeryLidsey}.

The quantity of interest is $\zeta^\subt_\sigma$, the tachyon curvature perturbation.
Beyond linear order in perturbation theory there are nonadiabatic pressures in the model
which will source the time evolution of $\zeta^\subt$ on large scales.  The contribution
$\zeta^\subt_\sigma$ is the term which is amplified during the preheating phase and which
will come to dominate $\zeta^\subt$ at late times.  We therefore focus on $\zeta^\subt_\sigma$,
since any significant nongaussianity will arise due to this term.

One of the principal results of \cite{BC} was the computation of the
tachyonic contribution to the second order tachyon curvature perturbation in terms of
the first order tachyon fluctutations $\done\sigma$:
\begin{eqnarray}
  \zeta^\subt_{\sigma} &\cong& \frac{\kappa^2}{\epsilon}
\int_{\tau_i}^{\tau}d\tau' \left[
                               \frac{\left(\done\sigma'\right)^2}{\sH(\tau')} \right. \nonumber 
\\ &-& \left. \frac{\sH(\tau')^2}{\sH(\tau)^3}\left( \left(\done\sigma'\right)^2
- a^2 m_{\sigma}^2\left(\done\sigma\right)^2\right) \right]
\label{final}
\end{eqnarray}
where $\kappa^2 = \Mpl^{-2} = 8\pi G_{N}$, $\epsilon$ is the slow
roll parameter, $\epsilon=\frac12 M_p^2 (V'(\varphi)/V)^2$, $\tau$ is the
conformal time, $\tau = -[H a (1-\epsilon)]^{-1}$, ${\cal H}$ is the
conformal time Hubble parameter, ${\cal H} = 1[\tau(1-\epsilon)]^{-1}$,
and all factors in the integrand are evaluted at $\tau'$ unless
otherwise indicated.  In deriving (\ref{final}), we have performed
partial integrations in which surface terms at the initial time
were dropped; hence (\ref{final}) is only valid for perturbations
which are dominated by the tachyonic growth at late times.  For
such perturbations, there is little sensitivity to the value taken for
$\tau_i$.  An analogous result was derived for hybrid inflation (not considering
preheating) in \cite{EV}.
Metric perturbations during preheating have also been discussed in
\cite{firstorderreheating}-\cite{FB2}.

Since $\zeta^\subt_\sigma$ depends only on the first order tachyon fluctuation $\done\sigma$,
and not on $\dtwo\sigma$ \footnote{Indeed, as was shown in \cite{BC}, $\dtwo\sigma$ decouples 
from the gauge invariant quantity up to second order in perturbation theory.} we can drop
the superscript $(1)$ and denote $\done\sigma$ by $\delta\sigma$ in subsequent text.

The expression (\ref{final}) satisfies an important consistency check,
namely that it is a local expression.  Nonlocal
operators $\triangle^{-n}$, powers of the inverse Laplacian, arise at intermediate steps in
the calculation, using the generalized longitudinal gauge,
which separates metric perturbations into scalar, vector and tensor
components.  In the process of decoupling these to solve for the 
curvature perturbation, one must apply $\Linv$.  The nonlocal terms
should cancel out of physical quantities, similarly to
electrodynamics in Coulomb gauge.   The second order curvature
perturbation  is related to the observable CMB temperature
fluctuations in a nontrivial way, so this by itself does not prove
that $\zeta^\subt$ must be local.  However \cite{LangloisVernizzi}
has recently shown that under the conditions present in our model,
$\zeta^\subt$ should indeed by local.

Using (\ref{final}), it is possible to compute tachyonic 
contributions to the spectrum and bispectrum of the curvature
perturbation,
\beqa
\label{spectrum} 
 \left\langle \zeta^{(2)}_{k_1}\, \zeta^{(2)}_{k_2}\,
\right\rangle &\equiv& \delta(\vec k_1 + \vec k_2) \, S(k_i)\\
\label{bispectrum} 
\left\langle \zeta^{(2)}_{k_1}\, \zeta^{(2)}_{k_2}\,
 \zeta^{(2)}_{k_3} \right\rangle 
&\equiv& {\delta^{(3)}(\vec k_1 + \vec k_2 + \vec k_3)\over(2\pi)^{3/2}}\, \,
B(\vec k_1, \vec k_2 ,\vec k_3)\nonumber\\
\eeqa

\subsection{Tachyon mode functions}

To compute the correlations in (\ref{spectrum})-(\ref{bispectrum}),
we express the tachyon fluctuation in terms of creation and
annihilation operators, 
\beq
\label{quantum}
  \delta \sigma(\vec x,N) = \int {d^{\,3}k\over (2\pi)^{3/2}}
	\,a_k \, \xi_k(N)\, e^{i\vec k\cdot\vec x}+ {\rm h.c.}
\eeq
where the mode functions obey the linearized tachyon equation of
motion.  To make this equation more tractable, we have approximated
the tachyon mass dependence on  time ($N$) as being linear,
$m_\sigma^2 \cong -c H^2 N$, so that 
\begin{equation}
\label{mode}
  \frac{d^2}{dN^2}\delta\sigma_k + 3 \frac{d}{dN}\delta\sigma_k + 
  \left[\hat{k}^2e^{-2N} - c N\right]\delta\sigma_k = 0
\end{equation}
where $\hat{k} = k / H$.  We found that this technical assumption is
nearly always satisfied for model parameter values consistent with the
near scale-invariance of the CMB fluctuations; furthermore it is
always satisfied for parameters which lead to large nongaussianity.
The term $cN$ is proportional to the first term in the Taylor series 
for $e^{2\eta N} - 1$, where
$\eta \cong 4 M_p^2 m^2_\phi/ (\lambda v^4)$ is the usual slow roll
parameter for $\varphi$.  We thus demand that 
 $2\eta|N|\ll 1$ throughout inflation.  Notice that inflation is
ended by the tachyonic instability, not by the failure of the slow
roll conditions.
The coefficient $c$ is given by $c = 2\eta\lambda v^2/H^2$,  and $H^2 = V/(3M_p^2)$, with $V\cong
\frac14\lambda v^4$.  Using the COBE normalization
$V / (\Mpl^4 \epsilon) = 6\times 10^{-7}$ to eliminate the inflaton 
mass $m_\varphi$, we find that $c= 2.2\times 10^4 g \, M_p/v$.

Although eq.\ (\ref{mode}) has exact solutions in terms of Airy
functions when $k=0$, for general $k$ no closed-form solutions exist.
We therefore constructed solutions, using the WKB approximation,
or alternatively 
the adiabatic approximation.  The WKB approximation is valid in the
limit of $N\to\pm\infty$; these solutions are matched to each other
at $N_k$, a $k$-dependent intermediate value of $N$.  They have the
form
\beq
\label{solns}
	\xi_k \cong \left\{  \begin{array}{ll}
	\left({2 H\hat k^3}\right)^{-1/2} 
	\left(1 + i \hat k e^{-N}\right), & N < N_k \\
	b_k\,{ e^{-\frac32 N + \frac{9}{4c}z^{3/2}}(1+|z|)^{-1/4} }, &  N > N_k
\end{array} \right.
\eeq
with
\beq
\label{bk}
	b_k = {1-i\sqrt{c|N_k|}\over \sqrt{2 H}(c|N_k|)^{3/4}}
	{(1 + |z_k|)^{1/4} \over 
	\exp\left({9\over 4c}z_k^{3/2}\right)}
\eeq
where $z \equiv \left(1+\frac49 cN\right)$, 
$z_k \equiv \left(1+\frac49 cN_k\right)$, and the dividing time
between the small- and large-$N$ behavior for a given mode is
implicitly defined
by $N_k = \ln(\hat k/\sqrt{c}) - \ln\sqrt{|N_k|}$.  The matching time $N_k$ is
discussed in some detail in appendix A.  The alternative
method, using the adiabatic approximation, will be reviewed in 
section \ref{III}.  See also \cite{falsevacuumdecay} for a discussion of the
solutions of the mode function equation.

The solutions of (\ref{mode}) have been discussed in great detail in
\cite{BC}. However, a few comments are in order about the solutions
(\ref{solns}).  At early times $N < N_k$ the gradient term in the
Klein-Gordon equation dominates over the mass term,  $k^2/a^2 >
|m_\sigma^2|$, and the resulting mode functions look just like the
solutions for a massless field in de Sitter space.  These ultraviolet
modes are redshifted by the expansion of the universe into the
instability band where the mass term dominates the dynamics
$|m_\sigma^2| > k^2/a^2$.  (Because of the time dependence of
$m_\sigma$ the modes may reenter the massless regime for a brief
period of time; we have verified that this does not alter any of our
results.)  In this infrared regime (where the mass term dominates)
the Airy function solutions are appropriate.  We have checked the
solutions (\ref{solns}) against numerical solutions of (\ref{mode}),
and found good agreement.

\subsection{Integrated results}

Using the solution (\ref{solns}) in (\ref{final}), and going to
the limit of vanishing wave numbers, 
$\zeta^{(2)}_\sigma$ takes the form
\beqa
\label{timeint}
	\zeta^{(2)}_{k=0}(N_*) &=& {\kappa^2\over\epsilon}
	\int {d^3 p \over(2\pi)^{3/2}} 
	\left(\,a_p\, b_p\, + a^\dagger_p\, b_{-p}\right)^2
\nonumber\\&& 
\times\int_{{\rm max}(N_p,N_i)}^{N_*} f(c,N,N_*)\, dN 
\eeqa
where $f(c,N,N_*)$ is given by
\beqa
\label{feqn}
 && f(c,N,N_{\ast}) = 
{e^{-3N + \frac{9}{2c}z^{3/2}}
\left(1 + |z|\right)^{-1/2}}\times\qquad\nonumber\\
 && \qquad\left[ \frac{9}{4}\left( 1-e^{3(N-N_{\ast})}\right)
     \left|\sqrt{z} - 1 - \frac{2 c\,\, {\rm sign}(z)}{27(1+|z|)} 
	\right|^2 - \right. \nonumber\\
&& \left. \qquad \phantom{\left(\frac94\right)^2_2} 
c N e^{3(N-N_{\ast})} \right]
\eeqa
It should be noted that the dominant time-dependence is determined
by the combination $-3N + \frac{9}{2c}z^{3/2}$ in the exponent.
The $e^{-3N}$ decay factor is typical of massive modes, 
which redshift as $\delta\sigma_k\sim a^{-3/2}$.  The
$e^{\frac{9}{2c}z^{3/2}}$ growth factor is a result of the tachyonic
instability.  As we noted in the discussion following eq.\ (\ref{final}),
(\ref{timeint}) is valid only when the late-time behavior dominates.
Therefore an important consistency condition for all of our analysis
is that
\beq
\label{condition}
\frac{9}{2c}z_*^{3/2} \equiv 
\frac{9}{2c}\left(1+\frac49 c N_*\right)^{3/2} > 3|N_i|\, .
\eeq
If this condition is violated then the preheating is not playing any role
in the dynamics and we can safely assume that no significant nongaussianity
is produced.

Using (\ref{feqn}), the correlation functions which give the spectrum and bispectrum 
from (\ref{spectrum}) and (\ref{bispectrum}) can then be computed
as
\begin{equation}
\label{Seq}
  S = 2{\kappa^4\over\epsilon^2}\int {d^{\,3}p\over (2\pi)^3} |b_p|^4
\left[\int_{\mathrm{max}(N_p, N_i)}^{N_*} dN\, f(c,N,N_*)\right]^2
\end{equation}
and
\beq
\label{Beq}
	B = 8{\kappa^6\over\epsilon^3}\int {d^{\,3}p\over (2\pi)^3} |b_p|^6\left
[\int_{\mathrm{max}(N_p, N_i)}^{N_*}\!\!\!\!\!\!\!\! dN\, f(c,N,N_*)\right]^3
\eeq

The tachyonic contribution to the spectrum cannot exceed the
experimentally inferred inflaton power spectrum, $\sqrt{P_\varphi(k)}
\cong 2\pi \times 10^{-5} k^{-3/2}$, leading to the bound 
$S<P_\varphi$.  Moreover the bispectrum is related to the
nonlinearity parameter $f_{NL}$ by  $B =
-\frac{6}{5}f_{NL}\left(P_\varphi(k_1)P_\varphi(k_2) + 
\mathrm{perms} \right)$, which at equal momenta $k_i$ leads to
\beq
f_{NL} =-\frac{5}{18}\,{B\over P_\phi^2}
\label{fnl}
\eeq
The current experimental constraint is $|f_{NL}|\lsim 100$.  
By analogy, we also define a parameter $f_L$ for the spectrum
as
\beq
	f_L = {S\over P_\phi}
\label{fl}
\eeq
and demand that $|f_L| < 1$.\footnote{One might consider being more conservative and imposing, 
say, $|f_L| < 0.01$, rather than $|f_L| < 1$, as we have done.  Because the effect turns
on exponentially fast, our exclusion plots are actually quite insensitive to the value assumed
for $f_{L}$ and $f_{NL}$.  For example, the exclusion plots for $|f_{NL}| < 1$ are visually
hard to distinguish from those for $|f_{NL}| < 100$.}

\section{Constraints on hybrid inflation parameter space}
\label{III}

By numerically evaluating the integrals (\ref{Seq}) and (\ref{Beq})
and applying the experimental limits on the inflaton power spectrum
and bispectrum, we were able to find constraints on the hybrid
inflation parameter space.   We update these bounds in the present
section.

\subsection{The issue of scale invariance}

In \cite{BC} we noted that it is possible for the tachyonic
contributions to the spectrum or bispectrum either to be  nearly
scale invariant ($S\sim 1/k^3$, $B\sim 1/k^6$), or else to  badly
violate scale invariance ($S$, $B\sim k^0$).  In the latter case, the
spectral index for the tachyon contribution to the two-point  function is
$n=4$.  However we did not clearly differentiate between these two
regimes in the limits presented in \cite{BC}, an omission which we
rectify here.  

The two regimes, scale-invariant and nonscale-invariant, can be 
understood in reference to the condition (\ref{condition}) which
must be satisfied in order for tachyonic preheating to play any
significant role.  There are two ways to satisfy
(\ref{condition}).  One is to take $c N_\star \gg 1$, which
usually requires $c > 1$.  This is the regime in which the
tachyon mass is not small compared to $H$ during most of
inflation, and so it corresponds to  nonscale-invariant
fluctuations of $\delta\sigma$.  The tachyon fluctuations are
Hubble damped as $\delta\sigma \sim a(t)^{-3/2}$ prior to
inflation, but this suppression can be overcome on large
scales  if the amplification during the preheating phase is 
sufficiently large, which typically requires  very small values
of the self-coupling $\lambda \ll 1$.   This nonscale-invariant
regime corresponds to  a region of the parameter space where the
waterfall condition of hybrid inflation is satisfied.

The second way to satisfy (\ref{condition}) is to take $c|N| < 1$
for all  $N \in \left[N_i,N_\star\right]$.  This gives a
scale-invariant spectrum for the tachyon and also for
$\zeta^\subt_\sigma$, which is most easily seen by writing the
tachyon  mass-squared as ${|m_\sigma^2|}/{H^2} = c |N|$.
It is clear that if $c|N| < 1$ for all $N \in
\left[N_i,N_\star\right]$ then the tachyon field will have been
light compared to the Hubble scale for all $\sim 60$ e-foldings
of inflation which ensures a nearly scale-invariant spectrum for
the tachyon.  Also, in this case the  instability will typically
take several e-foldings to complete so that $N_\star > 1$.   This
scale-invariant regime corresponds to a region of the parameter
space where the usual  waterfall condition of hybrid inflation is
violated.

\subsection{Nonscale-invariant case}

\begin{figure}[htbp]
\bigskip \centerline{\epsfxsize=0.5\textwidth\epsfbox{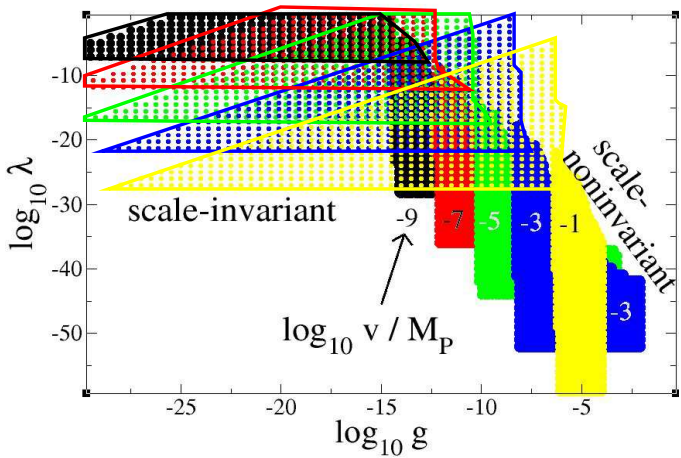}}
\caption{Excluded regions of the hybrid inflation parameter
space, for $\log_{10}v/M_p = -1,-3,-5,-7,-9$.
}
\label{figN}
\end{figure}

In the nonscale-invariant case $f_{L}$ and $f_{NL}$ depend on $k$.  Because the 
tachyon spectrum
is blue in this case the strongest constraint comes from evaluating $f_{L}$, 
$f_{NL}$ at the
largest values of $k$ which are measured by the CMB.  In deriving our 
constraints we 
conservatively take this to be $k = e^6 H e^{-N_e}$ where $N_e$ is the 
total number of
e-foldings of inflation (\ref{Ne2}).  The resulting constraints
in the plane of $\log_{10} g$ and $\log_{10}\lambda$
are shown in the right-hand region of figure \ref{figN}, for
several values of $\log_{10} v/ M_p$.  We find that the most
stringent constraints come from $f_L$ rather than $f_{NL}$, so 
we expect to see distortions of the spectrum rather than
nongaussianity at the left-hand boundaries of the excluded regions.
(The other boundaries are unexcluded for the different reasons
described in the next paragraph.)  In
comparing these excluded regions to figure 12 of \cite{BC}, one
sees that they are smaller than in our previous work, in the upper
right-hand corner.  This is due
to correcting an error in \cite{BC}, in which we failed to apply
the condition (\ref{condition}) restricting the validity of our analysis.

In computing $f_{L}$, $f_{NL}$ over a wide range of
$g,\lambda,v/\Mpl$ we also checked that the additional
assumptions were respected: the tachyon mass-squared $m_\sigma^2$
varies linearly with the number of e-foldings, which was shown in
\cite{BC} to require $g {v}/{\Mpl} < 10^{-5}$; the false vacuum
energy density  $\lambda v^4 /4$ dominates  during inflation,
leading to the bound $g > 460 \lambda
\left({v}/{\Mpl}\right)^3$;  the reheat temperature exceeds 100
GeV, so that baryogenesis can occur at least during the
electroweak phase transition, leading to the lower bounds on
$\lambda$. 

\subsection{Scale-invariant case; the adiabatic approximation}

On the left-hand side of figure \ref{figN}, we display new
constraints for which the spectrum and bispectrum are
scale-invariant.  In contrast to the right-hand side, $f_{NL}$
provides the dominant constraint here, so that nongaussianity
is playing the important role.  To obtain these results,
we employed a different approximation for the tachyon mode
functions, namely the adiabatic approximation, described in
appendix F of \cite{BC}.  Because the tachyon has a small mass
during the entirety of inflation (subsequent to horizon crossing),
its mass is changing slowly, and we can use the standard mode
functions for light fields, but with a time-dependent mass:
\beq
\label{appquantum}
  \delta \sigma(x) = \int {d^{\,3}k\over (2\pi)^{3/2}}
{H\over\sqrt{2k^3}} (-k\tau)^{\eta_\sigma(\tau)}
	e^{ikx}\,a_k + {\rm h.c.}
\eeq
Here $\eta_\sigma = M_p^2 V_{,\sigma\sigma}/V$ is the slow-roll
parameter for the tachyon, given by
\beq
\label{etas2}	
	{\eta_\sigma}(\tau) = 8\eta\, {M_p^2\over v^2}\,
	\ln| H\tau|
\eeq
where $\eta = M_p^2 V_{,\varphi\varphi}/V$.  We have also verified that the 
solution
(\ref{appquantum}) can be reproduced by (\ref{solns}) in the appropriate limit.

Since the mode
functions have a simple form in the adiabatic approximation, it is possible to 
go farther
analytically in this case.  Notably, we could find an implicit
equation for $N_*$ after evaluating the integral
(\ref{end_of_inflation}):
\beq
\label{nstar}
	N_* \cong { v/ M_p\over 15000\, N_e\,
g} \ln\left[ 1 + 2\times 10^6\,{N_*g}
\left({M_p\over v}\right)^{3}\right]
\eeq
This expression for $N_*$ is much easier to evaluate than the
one which arises in the WKB approximation since the latter
leads to a numerical integral $f(N; \lambda, g,v)$
which must be inverted to find the $N_*$ which satisfies
$f(N_*; \lambda, g,v)=v^2/4$.

Moreover, the time ($N$) integral in (\ref{final}) can be
evaluated explicitly using the saddle point method, since it is
dominated by the exponential growth near $N=N_*$.  Namely, an
integral of the form $\int dN e^g$ is approximated by
$e^{g_*}/\sqrt{|g'_*|}$ where $g_*$ is  the maximum value (at
$N=N_*$) and $g'_*$ is the derivative evaluated at the same 
point.  Defining $\eta_f$ to be $|\eta_\sigma|$ at $N=N_*$, this
results in the expression 
\beqa
	B(\vec k_i) &=& 4H^{6\eta_f} d_*^3\int
	{d^{\,3}p\over (2\pi)^3}\,{|p|^{-3-2\eta_f}\,
	|p-k_3|^{-3-2\eta_f}}\nonumber\\
	&&\left({|p+k_2|^{-3-2\eta_f}} + 
	{|p+k_1|^{-3-2\eta_f}}\right)
\eeqa
for the bispectrum,\footnote{In \cite{BC}, the conformal time
when the instability starts is (perhaps confusingly named) 
$\tau_* = -1/H$, due to our choice of $N=0$ for the beginning
of the instability.}  where 
$d_* = H^2 \kappa^2 \eta_f  e^{2\eta_f N_*}/(2\epsilon)$.  In the
limit of small $c\sim \eta_f$, this is manifestly nearly scale invariant,
$B\sim 1/k_i^{6(1+\eta_f)}$, by power-counting the divergent behavior of
the integral in the infrared.  The divergence must be cut off
in the usual way, by ignoring modes with $p$ smaller than the
horizon.  Numerically evaluating the remaining $p$ integral for
 $k_i\sim k$, and using $k\tau_*\sim e^{N_i}$ for modes
near the horizon,\footnote{The horizon-crossing condition is
$k\tau_i=1$, and $\tau_*/\tau_i = e^{N_i}/e^0$.}
we find 
\beq
	B(k)\cong 45\, k_i^{-6(1+\eta_f)}\left({\kappa^2 H^2\eta_f
	e^{2\eta_f N_e}\over 2\pi \epsilon}\right)^3
\eeq
Further using the
COBE normalization to write $\kappa^2 H^2/\epsilon = 2\times
10^{-7}$, we find that the nonlinearity parameter is
\beq
	f_{NL} = -2.6\times 10^{-5} \left(\eta_f e^{2\eta_f N_e}
	\right)^3.
\eeq
Moreover the COBE normalization implies $\eta_f = 7360 N_* g
M_p/v$.  Demanding that $|f_{NL}|<100$ gives the new excluded
regions on the left-hand side where $\xi_k$ is the mode function for the fluctuations (discussed
below).  This happens at some time after the onset of the instability.
For a wide range of parameters (including the values originally considered
in \cite{hybrid}) one has $N_\star \ll 1$ so that the symmetry breaking
completes on a time scale short compared to the Hubble time.  This is the
usual waterfall condition regime of hybrid inflation.  In the present work
we will consider both the possibilities that $N_\star \ll 1$ an also 
$N_\star \gsim 1$.of figure \ref{figN}.  Unlike the
nonscale-invariant regions, these have nongaussianity being the
dominant effect, rather than the tachyon contribution to the
spectrum.

We have claimed that in the scale-invariant regime the dominant
constraint is coming from $f_{NL}$ and not from $f_L$, contrarily to
the nonscale-invariant regime.  We
now justify this claim.  Repeating the steps above for the tachyon spectrum,
$S$, one obtains
\begin{equation}
  |f_L| \sim 10^{-6}\left(\eta_f e^{2\eta_f N_e}
	\right)^2
\end{equation}
Thus, in the scale-invariant regime, the linearity and nonlinearity parameters
are related as
\begin{equation}
\label{nonlinearity_vs_linearity}
  |f_{NL}| \sim 10^{6} |f_L|^{3/2}
\end{equation}
so that $|f_{NL}| > |f_L|$ except when $|f_L|$ is extremely small.  This
demonstrates that it is indeed possible to obtain significant nongaussianity
in this region of the parameter space.  We have verified that the result 
(\ref{nonlinearity_vs_linearity}) can also be derived using the mode function 
solutions (\ref{solns}).

\section{Implications for Brane-Antibrane Inflation}
\label{IV}

We now apply our results to a popular model of inflation from  string
theory, brane-antibrane inflation, correcting and extending our
preliminary results in this direction in \cite{BC}. This can be done
by mapping the low-energy effective action for the brane-antibrane
system onto the hybrid inflation model.  We focus on the popular
KKLMMT scenario \cite{KKLMMT,realistic,tye}  which reconciles brane
inflation with modulus stabilization  using warped geometries with
background fluxes for type IIB string theory vacua  \cite{vacua}.  In
this model, the antibrane is at the bottom of a Klebanov-Strassler
(KS) throat \cite{KS}, with warp factor $a_i\ll 1$, and the brane
moves down the throat.\footnote{See \cite{braneNG} for other discussions on
nongaussianity in string theory models of inflation.}\ 
Within the KS throat the geometry is well approximated by
\[
  ds^2  = a(y)^2 g_{\mu\nu}dx^{\mu}dx^{\nu} + dy^2 + y^2 d\Omega_5^2
\]
where $y$ is the distance along the throat, $a(y)\cong e^{ky}$ is the warp
factor and $d\Omega_5^2$ is the metric on the base space of the corresponding
conifold sigularity of the underlying Calabi-Yau space.  In the subsequent analysis
we ignore the base space and treat the geometry as $AdS_5$.

In the following we compute only the nongaussianity which is due to
the preheating dynamics and ignore the possible effects of
the inflaton sound speed  \cite{DBI,leblond,braneCs}; hence our results
may be thought of as a lower bound on the nongaussianity from brane
inflation.

In string theory the open string tachyon $T$ between a 
D$3$-brane and antibrane,\footnote{We restrict ourselves to
inflation models driven by D$3$-branes since inflation driven by
higher dimensional branes have problems with overclosure of the
universe by defect  formation \cite{BBCS}.} separated by
a distance $y$, is described by the action \cite{Sen}
\begin{eqnarray}
  S_{\mathrm{tac}} &=& -\int d^{4}x \,  \sqrt{-g} \, \mathcal{L} \nonumber \\
  \mathcal{L} &=& V(T,y) \, \sqrt{1 + (a_iM_s)^{-2} 
g^{\mu\nu}\partial_{\mu}T^*\partial_{\nu}T}
  \label{DBI}
\end{eqnarray}
Here the small-$|T|$ expansion of the potential is
\begin{eqnarray}
  V(T,y) &=& 2\, a_i^4\,\tau_3 \left[ 1 + 
           \frac{1}{2} \left(\left(\frac{M_s y}{2\pi}\right)^2
 - \frac{1}{2}\right)|T|^2 \right. \nonumber\\
         && \,\,\,\,\,\,\,\,\, + \left. \mathcal{O}(|T|^4) + \cdots 
	\phantom{\frac{(M_s y)^2}{(2\pi)^2}}
\!\!\!\!\!\!\!\!\!\!\!\!\!\!\!\!\!\!\!\right].
\label{smallT}
\end{eqnarray}
where $M_s$ is the string mass scale, $\tau_3 =
g_s^{-1}M_s^4/(2\pi)^3$ is the D3-brane 
tension, and $g_s$ is
the string coupling.  
Notice that in the warped throat scenario the instability does not 
set in until the 
branes are separated by the unwarped string
length,\footnote{There is some confusion on the literature on this
point, with some papers having stated that the instability is
determined by the warped string length, but this is not the case \cite{leblond}.
We thank L.\ Leblond for pointing this out to us.}  
$(a_i M_s)^{-1}$.
An interesting difference between the string tachyon and that of
ordinary hybrid inflation is that (at $y=0$) the tachyon potential $V(|T|)$
in the string case does not have a local minimum; rather
\beq
  V(T,0) \sim \tau_3 \, e^{-|T|^2 /4}
\label{exp_pot}
\eeq
The potential is minimized as 
$T\to\infty$.  Therefore $T$ does not have a VEV.  Nevertheless,
the unstable brane-antibrane system decays into closed strings
soon after the instability begins, and the large-$T$ part of the
potential is not meaningful for determining the actual evolution
of the tachyon.  In hybrid inflation, it is also true that the
end of inflation occurs somewhat before the fluctuations of the
tachyon become as large as the VEV.  We will see that even in the
absence of a $T^4$ coupling, we can still define the equivalent
of $\lambda$  and $v$ for the brane-antibrane system, by equating 
$\frac14\lambda v^4$ to the false vacuum energy, and $\lambda v^2$ to the
tachyon mass scale.  This amounts to replacing the condition 
for the end of inflation (\ref{end_of_inflation}) by 
\begin{equation}
\label{eoi2}
  \left. \int \frac{d^3k}{(2\pi)^3} |\xi_k|^2 \right|_{N = N_\star} = 
{\hbox{false vacuum energy}\over|\hbox{tachyon mass}|^2}
\end{equation}
Despite that fact that the tachyon potential is only minimized at
$|T| \rightarrow \infty$ the condition (\ref{eoi2}) is quite
reasonable. Detailed numerical simulations of the symmetry breaking
in the theory (\ref{DBI}) were performed in \cite{BBCS}.  Comparing
to the analysis of \cite{BBCS} one finds that $N_\star$ as defined in
(\ref{eoi2})  roughly corresponds to the time at which
singularities in the spatial gradients of  the tachyon field 
form \cite{singularity1}.  The appearance of 
singularities  within a finite time 
corresponds to the formation of lower dimensional
branes \cite{singularity2} and  hence by $N=N_\star$ the inflaton
field ceases to exist as a physical degree of freedom.  This
means that, as in our previous analysis, for $N > N_\star$ there no
longer exists any nonadiabatic pressure (since only one field, the
tachyon, is dynamical) and the large scale curvature perturbation
becomes conserved to all orders in perturbation  theory
\cite{conserved}.

The effective values of the couplings
can be found by rewriting the action  in
terms of the canonically normalized fields $\sigma = a_i\sqrt{\tau_3} T
/ M_s$ and $\varphi = \sqrt{\tau_3} y$ 
(see equations 3.6, 3.10 or C.1 in \cite{KKLMMT}), and then matching 
to the hybrid inflation potential (\ref{pot}).
This gives the correspondence
\begin{eqnarray}
  v &=& \sqrt{\frac{2}{\pi^{3}}}\frac{a_i M_s}{\sqrt{g_s}} 
\label{bi_v} \\
  \lambda &=& \frac{\pi^3}{4} g_s \label{bi_lambda} \\
  g &=& \sqrt{2\pi g_s}\, a_i \label{bi_g}
\end{eqnarray}

For the analysis of the preceding sections to be valid,
the inflaton potential must be well-described by 
$V_0 + \frac12 m_\varphi^2\varphi^2$ during
the relevant e-foldings of inflation.  
The full potential can be written as
\begin{equation}
\label{inf_pot}
  V_{\mathrm{inf}} = \frac{m_\varphi^2}{2}\varphi^2 
  + V_0\left(1-\frac{\nu}{4\pi^2}\frac{V_0}{\varphi^4}\right)
\end{equation}
where $V_0 = 2 a_i^4 \tau_3$
and $\nu$ is a geometrical factor which 
is given $\nu = 27 / 16$ for the KS throat.
It is typical to parameterize the inflaton mass in terms of the dimensionless quantity
$\beta$ as $m_\varphi^2 = \beta H_0^2$ where
$H_0^2 = V_0/({3\Mpl^2})$.  Using the COBE normalization, we find
that 
\beq
	\beta = 10^{7/2}\, a_i^3\,\left(M_s\over M_p\right)^3.
\eeq
Demanding that the mass term in (\ref{inf_pot}) dominate over the
Coulomb term even
when the branes are separated by the local string length yields a lower bound on 
$\beta$:
\[
  \beta > 324 \pi^4 g_s^2 a_i^{10} \left(\frac{\Mpl}{M_s}\right)^2
\]
The parameter $\beta$ is also bounded from above by the requirement that 
$|n-1| \lsim 10^{-1}$ which corresponds to 
$g_s^2 a_i^{10}({\Mpl}/{M_s})^2 \ll 5\times10^{-6}$.

Our results apply only in the case that $\beta > 0$; moreover
the case where $\beta<0$ does not make sense from the string
theory point of view, since $\varphi=0$ denotes the bottom of the
throat, and the brane must roll toward that point, not away from
it \cite{BCDF}.


Our preliminary results about this in \cite{BC} suffered from the
neglect of the condition (\ref{condition}); moreover we unduly
restricted the full string parameter space by assuming that the scale
of inflation was determined by the COBE normalization; that is not
the case.  As in hybrid inflation, we still have three parameters
even after normalizing the spectrum, which we can take to be $g_s$,
$a_i$ and the ratio of the warped string scale to the Planck scale, 
$a_i M_s/ M_p$.  Taking into account the additional restrictions on
$\beta$, we find that the scale-noninvariant exclusions (right-hand
side of fig.\ \ref{figN}) do not survive at all in the KKLMMT model;
however all the scale-invariant ones do. Therefore this model has the
potential for producing large nongaussianity, and is even constrained
by producing too high levels of nongaussianity. 

\begin{figure}[htbp]
\bigskip \centerline{\epsfxsize=0.5\textwidth\epsfbox{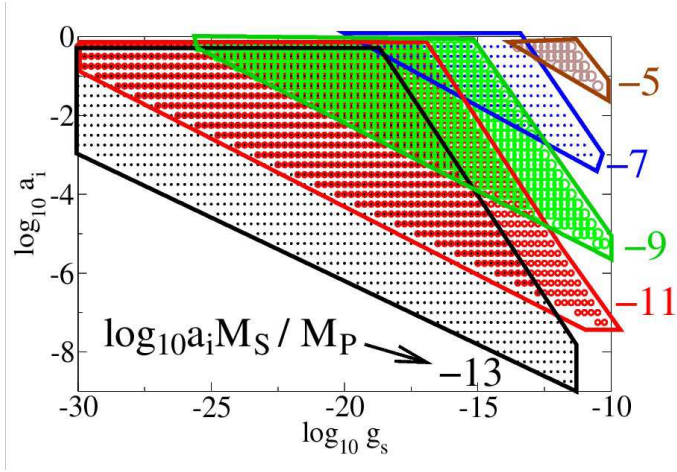}}
\caption{Excluded regions of the KKLMMT brane-antibrane inflation
 parameter space, in the plane of $\log_{10} a_i$ versus $\log_{10} g_s$
for $\log_{10}a_i M_s /M_p =-13,-11,\dots,-5$
from nongaussianity.
}
\label{string-ex1}
\end{figure}

The constraints in the string parameter space are shown in figure
\ref{string-ex1}. The excluded regions shown here correspond to very
small values of $g_s\lsim 10^{-10}$.  In the simplest way of
connecting type IIB string theory to low-energy phenomenology, the
gauge couplings of the Standard Model are related to $g_s$ by running
down from the string scale, which would render such small values of
$g_s$ incompatible with observations.  However, type IIB strings are
dual to themselves under SL(2,Z) transformations which take $g_s\to
1/g_s$.  In the dual picture, the string coupling is very large, and
the gauge dynamics at the string scale would be confining.  It is 
conceivable that the Standard Model arises as a remnant of a strongly
coupled gauge theory at the high scale, similar to technicolor models.
In this case, the small values of $g_s$ which give rise to large
nongaussianity could still be compatible with particle physics
constraints.  

\section{Implications for P-term Inflation}
\label{V}
In realistic models of hybrid inflation from supergravity, the 
potential is generated by F- or D-terms.  P-term inflation is a
class of models which combines the two kinds of terms and can
interpolate between them \cite{Pterm}.  
The potential for P-term inflation, along the inflationary trajectory,
 is
\begin{equation}
\label{p_pot}
  V_{\mathrm{inf}} = \frac{g^2 \xi^2}{2}\left(1 
  + \frac{g^2}{8\pi^2}\ln\frac{\varphi^2}{\varphi_c^2}
  + \frac{f}{8}\frac{\varphi^4}{\Mpl^4} + \cdots \right)
\end{equation}
where $\cdots$ denotes terms of order $\varphi^6/\Mpl^6$ and higher.
Here $\varphi_c=\xi =\sqrt{|\xi_+|^2+ \xi_3^2}$ is defined in terms of
 two
Fayet-Iliopoulos parameters $\xi_+$ and $\xi_3$, and 
  in (\ref{p_pot})
$f$  must lie in the interval  $0 \leq f \leq 1$, since it is defined
as $f = |\xi_+|^2/ \xi^2$. The limits $f=0$ and $1$ correspond to
D-term \cite{Dterm} and  F-term \cite{Fterm} models, respectively. 
We will consider each of these limits separately.  We do not consider
the complications which arise when these models are coupled to moduli
fields \cite{ACDavis}.

As in the models previously considered the false vacuum energy dominates 
during inflation and the Hubble scale is given by
\begin{equation}
\label{p_H}
  H \cong \frac{g \,\xi}{\sqrt{6}\Mpl}
\end{equation}
The inflaton couples to two scalar fields $\Phi_\pm$, of which
one linear combination $\sigma$ is tachyonic.  Its 
mass-squared is given by
\begin{equation}
\label{p_m_sigma}
  m_\sigma^2 = g^2\left(\varphi^2 - \xi\right)
\end{equation}
By comparing (\ref{p_H}) and (\ref{p_m_sigma}) to the hybrid inflation potential 
(\ref{pot}) we can determine the hybrid inflation model parameters as
\begin{eqnarray}
  \lambda &=& \frac{g^2}{2} \label{p_lambda} \\
  v &=& \sqrt{2\xi} \label{v_lambda}
\end{eqnarray}
The coupling $g$ retains its original meaning in P-term inflation.

\subsection{D-term Inflation}

D-term inflation corresponds to taking $f=0$ in (\ref{p_pot}).
During a slow roll phase the inflaton field evolves as
\[
  \varphi_0(t)^2 = \varphi_c^2 - \frac{g^3 \xi}{2\sqrt{6}\pi^2}(t-t_c)
\]
which implies that $m_\sigma^2$ varies linearly with the number of e-foldings.  Scales
relevant for the CMB left the horizon when $\varphi = \varphi_N$ where
\[
  \varphi_N^2 = \varphi_c^2 + \frac{g^2 N}{2\pi^2}\Mpl^2 = \xi + \frac{g^2 N}{2\pi^2}\Mpl^2
\]
Two distinct regimes are possible depending on the value of the coupling $g$.
It is often assumed that $g$ is relatively large so that $g^2 N/(2\pi^2) \gg \xi/\Mpl^2$
\cite{Dterm2} which gives the correct amplitude of density perturbations with 
$\xi \cong 10^{-5} \Mpl^2$
and requires $g \gsim 2\times 10^{-3}$ for consistency.  In this regime 
$\varphi_N \gg \varphi_c$ so that slow roll at the onset of the instability is not guaranteed
and our previous analysis of the tachyon mode functions does not apply.  However, in this
regime it is also difficult to satisfy the constraints 
coming from the cosmic string tension, to avoid overproduction of
cosmic strings, $\xi \lsim 4\times 10^{-7}$ \cite{Dterm_strings}.
(Ref. \cite{loophole} has pointed out that the constraints on the cosmic
string tension can be weakened by incorporating the effect which strings have
on the observed spectral index.)

We are therefore driven to consider D-term inflation in the regime of
small coupling $g^2$ so that $g^2 N/(2\pi^2) \ll \xi/\Mpl^2$ and $\varphi_N
\cong \varphi_c$.  In this case we are guaranteed that the universe
will still be in a slow roll phase at the onset of the instability
and our previous analysis holds without modification.  This
corresponds to very small couplings $g \ll 2\times 10^{-3}$, however,
there is no obstruction to taking such a small coupling since $g^2$
is not necessarily related to the gauge coupling constant in a GUT \cite{Pterm}. 
In this regime the COBE normalization fixes $\xi \cong 7\times
10^{-4} g^{2/3} \Mpl^2$ so that $g$ is the only independent model
parameter.   The cosmic string constraint $\xi \lsim 4\times 10^{-7} \Mpl^2$
then restricts the coupling $g$ to be smaller than $g \lsim 1.3
\times 10^{-5}$.

Applying our previous analysis of hybrid inflation to D-term 
inflation,\footnote{See \cite{Dterm_preheat} for further discussion of preheating
in D-term inflation.}\ including the additional constraints mentioned above,
we find that there is a range of couplings,
\beq
  -10.0 <  \log_{10} g \lsim -8.7
\label{dterm-ex}
\eeq
which is ruled out because of the spectral distortion constraint,
in the scale-noninvariant region of figure \ref{figN}.
On the other hand there is no constraint coming from
nongaussianity for this model.

\subsection{F-term Inflation}

F-term inflation \cite{Fterm,Fterm2} corresponds to taking $f=1$ in (\ref{p_pot}).  In this 
case the
dynamics are somewhat more complicated than the D-term model.  Again there are
two possible regimes: a large coupling regime where $\varphi_N \gg \varphi_c$ and
our previous analysis does not apply and also a small coupling regime where 
$\varphi_N \cong \varphi_c$ and our previous analysis does apply.  The large coupling
regime corresponds to $g \gsim 2\times 10^{-3}$ and again the cosmic string tension
constraints are difficult to satisfy (see, however, \cite{enlarging}).  
We are therefore driven to consider only the
small coupling regime, $g \lsim 2\times 10^{-3}$.  For couplings 
$3\times 10^{-7} \ll g \lsim 2\times 10^{-3}$ it can be shown that the quadratic term 
$\varphi^4/\Mpl^4$ in the potential (\ref{p_pot}) can be neglected and the dynamics 
is identical to D-term inflation, which we have already considered.  Thus we consider
only the F-term model for $g \ll 3 \times 10^{-7}$ since this is the only region of
parameter space for which the model differs significantly from the D-term model.
\footnote{We have neglected the intermediate regime $0.06 \lsim g \lsim 0.15$
which will not yield significant nongaussianity or spectral distortion.}

For $g \lsim 3\times 10^{-7}$, so the $f$-term is dominating the
potential, the slow roll parameter $\epsilon = \xi^3/(8 M_p^6)$,
and the COBE normalization fixes
$\xi \cong 6.7\times 10^{6} g^2 \Mpl^2$
and the cosmic string tension is within observational bounds for $g
\lsim 2\times 10^{-7}$. Again applying the general hybrid inflation
constraints, we find the excluded region
\beq
  -13.0 \lsim \log_{10} g \lsim -9.5
\label{fterm-ex}
\eeq
which, as in the case of D-term inflation, comes from the 
$k^3$ spectral distortion effect rather than nongaussianity.

For more general P-term models with $0< f < 1$ we expect the
excluded regions to interpolate between (\ref{dterm-ex}) and
(\ref{fterm-ex}). In deriving our constraints we have been driven to
the small coupling regime by the requirement that the cosmic sting tension
be within observational bounds.  Our analysis does not give any significant
constraint on the string theoretic D3/D7 model \cite{D3/D7} since in this case the cosmic
strings are not stable and there is no motivation to consider the small values
of the coupling $g$ in (\ref{dterm-ex}) and (\ref{fterm-ex}).  Indeed, such small 
couplings are difficult to motivate from string theory \cite{D3/D7_small_g}.

\section{The Case of a Complex Tachyon}
\label{VI}

In the preceeding sections we have applied the results of \cite{BC},
which were derived under the assumption that $\sigma$ is a real
field, to models in which the tachyon is actually complex.  In doing
so we have assumed that the generalization of the analysis of 
\cite{BC} to the case of a complex tachyon does not significantly
modify the exclusion plot, figure \ref{figN}.  Here we verify this
claim.

\subsection{Cosmological Perturbation Theory for an $O(M)$ Multiplet}

Before restricting to the case of a complex tachyon we consider the somewhat
more general case of an $O(M)$ symmetric multiplet of tachyon fields $\sigma_A$ 
with $A = 1, \dots, M$.  The matter sector is expanded in perturbation theory as
\begin{eqnarray}
  \varphi(\tau,\vec{x}) &=& \varphi^\subz(\tau) + \done \varphi(\tau,\vec{x}) 
                            + \frac{1}{2}\dtwo \varphi(\tau,\vec{x}) \\
  \sigma_A(\tau,\vec{x}) &=& \done \sigma_A(\tau,\vec{x})
                           + \frac{1}{2} \dtwo \sigma_A(\tau,\vec{x}).
\end{eqnarray}
As in \cite{BC} the time-dependent vacuum expectation value (VEV) of the 
tachyon fields are set to zero $\langle \sigma_A \rangle \equiv \sigma^\subz = 0$ 
which is a consequence of the $O(M)$ symmetry of the theory.  Notice, however, that the 
tachyon field \emph{does} develop an effective VEV for the radial component
\[
  \langle |\sigma| \rangle \equiv \langle \sqrt{\sigma_A\sigma^A} \rangle
                           \not= 0
\]

We also assume that
\[
  \frac{\partial V}{\partial \sigma_A} = 
  \frac{\partial^2 V}{\partial \sigma_A \partial \varphi} = 0
\]
but $V$ is, for the time being, otherwise arbitrary.  Here and elsewhere the potential and 
its derivatives are understood to be evaluated on background values of the fields
so that $V = V(\varphi^\subz,\sigma^\subz)$, for example.

We consider only the $\dtwo G^0_0 = \kappa^2 \dtwo T^0_0$, 
$\partial_i \dtwo G^i_0 = \kappa^2 \partial_i \dtwo T^i_0$ and 
$\delta^i_j \dtwo G^j_i = \kappa^2 \delta^i_j \dtwo T^j_i$ 
equations since the second order vector and
tensor fluctuations decouple from this system.  In the case that $\sigma_A^\subz = 0$ 
the second 
order tachyon fluctations $\dtwo \sigma_A$ decouple from the inflaton and gravitational 
fluctuations up to second order and hence we do not need to solve for
$\dtwo \sigma_A$.  Note also that the Klein-Gordon equation for the inflaton fluctations
is not necessary to close the system.  In this section we 
sometimes insert the slow roll parameters $\epsilon$ and $\eta$ explicitly though we do not
yet assume that they are small.  We also introduce the shorthand notation 
$m_{\varphi}^2 \equiv \partial^2 V / \partial \varphi^2$.

The second order $(0,0)$ equation is
\begin{eqnarray}
&&  3 \sH \psi'^\subt  +(3-\epsilon)\sH^2 \phi^\subt - \Lap \psi^\subt \nonumber \\
&=& -\frac{\kappa^2}{2}  \left[ \varphi'_0 \dtwo \varphi' 
                      + a^2 \frac{\partial V}{\partial \varphi } \dtwo \varphi \right] 
+ \Upsilon_1
\label{00}
\end{eqnarray}
where $\Upsilon_1$ is constructed entirely from first order quantities.  Dividing 
$\Upsilon_1$ into inflaton and tachyon contributions we have
\[
  \Upsilon_1 = \Upsilon_1^{\varphi} + \Upsilon_1^{\sigma}
\]
where
\begin{eqnarray}
  \Upsilon_1^{\varphi} &=& 4(3-\epsilon)\sH^2\left(\phi^\subo\right)^2 
             + 2\kappa^2 \varphi'_0 \phi^\subo \done \varphi' \nonumber \\
             &-& \frac{\kappa^2}{2}\left( \done \varphi' \right)^2
             - \frac{\kappa^2}{2} a^2 m^2_{\varphi} \left( \done \varphi \right)^2 
             - \frac{\kappa^2}{2} \left(\Grad \done \varphi \right)^2 \nonumber \\
             &+& 8 \phi^\subo \Lap \phi^\subo
             + 3\left( \phi'^\subo \right)^2
             + 3\left( \Grad\phi^\subo\right)^2
\label{Upsilon1varphi}
\end{eqnarray}
and
\begin{eqnarray}
  \Upsilon_1^{\sigma} &=& - \frac{\kappa^2}{2} \left[ \done \sigma'_A \done\sigma'^A
  + \partial_i\done\sigma_A\partial^i\done\sigma^A \right. \nonumber \\ 
  &+& \left. a^2 \frac{\partial^2 V}{\partial\sigma_A\partial\sigma_B} 
      \done\sigma_A\done\sigma_B \right]. 
\label{Upsilon1sigma}
\end{eqnarray}
The divergence of the second order $(0,i)$ equation is
\begin{equation}
\label{0i}
  \Lap \left[ \psi'^\subt + \sH \phi^\subt \right] = 
  \frac{\kappa^2}{2} \varphi'_0 \Lap \dtwo \varphi + \Upsilon_2
\end{equation}
where $\Upsilon_2 = \Upsilon_2^{\varphi} + \Upsilon_2^{\sigma}$ is constructed entirely from 
first order quantities.  The inflaton part is
\begin{eqnarray}
  \Upsilon_2^{\varphi} 
             &=& 
2\kappa^2 \varphi'_0 \partial_i \left( \phi^\subo \partial^i \done \varphi \right)
             +
 \kappa^2 \partial_i \left( \done \varphi' \partial^i \done \varphi \right) \nonumber \\
             &-& 8 \partial_i \left( \phi^\subo \partial^i \phi'^\subo \right)
             - 2 \partial_i \left( \phi'^\subo \partial^i \phi^\subo \right)
\label{Upsilon2varphi}
\end{eqnarray}
and the tachyon part is
\begin{equation}
\label{Upsilon2sigma}
  \Upsilon_2^{\sigma} = 
  \kappa^2 \partial_i \left( \done \sigma'_A \partial^i \done \sigma^A \right).
\end{equation}
The trace of the second order $(i,j)$ equation is
\begin{eqnarray}
  && 3\psi''^\subt + \Lap \left[ \phi^\subt - \psi^\subt \right] 
  + 6 \sH \psi'^\subt \nonumber \\
  &+& 3 \sH \phi'^\subt + 3(3-\epsilon) \sH^2 \phi^\subt \nonumber \\
  &=& \frac{3\kappa^2}{2}\left[  \varphi'_0 \dtwo \varphi' 
      - a^2 \frac{\partial V}{\partial \varphi } \dtwo \varphi\right] + \Upsilon_3
\label{ij}
\end{eqnarray}
where $\Upsilon_3 = \Upsilon_3^{\varphi} + \Upsilon_3^{\sigma}$ is constructed 
entirely from first order quantities.  The inflaton part is
\begin{eqnarray}
  \Upsilon_3^{\varphi} &=& 12(3-\epsilon)\sH^2 \left( \phi^\subo\right)^2 
             - 6 \kappa^2 \varphi'_0 \phi^\subo \done\varphi' \nonumber \\
             &+& \frac{3\kappa^2}{2}\left(\done \varphi'\right)^2 
             - \frac{3 \kappa^2}{2} a^2 m^2_{\varphi} \left( \done \varphi \right)^2
             - \frac{\kappa^2}{2}\left(\Grad \done \varphi \right)^2 \nonumber \\
             &+& 3 \left( \phi'^\subo\right)^2 
             + 8 \phi^\subo \Lap \phi^\subo
             + 24 \sH \phi^\subo \phi'^\subo \nonumber \\
             &+& 7 \left( \Grad \phi^\subo \right)^2
\label{Upsilon3varphi}
\end{eqnarray}
and the tachyon part is
\begin{eqnarray}
  \Upsilon_3^{\sigma} &=& \kappa^2 \left[ \frac{3}{2}\done\sigma'_A\done\sigma'^A
        - \frac{1}{2} \partial_i\done\sigma_A\partial^i\done\sigma^A\right. \nonumber \\
   &-& \left. \frac{3}{2}a^2 \frac{\partial^2 V}{\partial\sigma_A\partial\sigma_B} 
       \done\sigma_A\done\sigma_B \right]
\label{Upsilon3sigma}
\end{eqnarray}

The derivation of the master equation which was presented in appendix B of \cite{BC} follows
here unmodified except for the new definitions of $\Upsilon_1^\sigma$, $\Upsilon_2^\sigma$
and $\Upsilon_3^\sigma$.  The master equation is
\begin{equation}
\label{master}
  \phi''^\subt + 2(\eta-\epsilon)\sH \phi'^\subt 
  + \left[ 2(\eta-2\epsilon)\sH^2 - \Lap \right] \phi^\subt = J
\end{equation}
where the source is
\begin{eqnarray}
  J &=& \Upsilon_1 - \Upsilon_3 + 4\Linv\Upsilon'_2 
  + 2(1-\epsilon+\eta)\sH \Linv\Upsilon_2 \nonumber \\ 
    &+& \Linv \gamma'' - (1 + 2\epsilon-2\eta)\sH\Linv\gamma'. \label{source}
\end{eqnarray}
and the quantity $\gamma$ is defined as
\begin{equation}
\label{gamma}
  \gamma = \Upsilon_3 - 3 \Linv \Upsilon'_2 - 6 \sH \Linv \Upsilon_2  
\end{equation}
We can split the source into tachyon and inflaton contributions 
$J = J^{\varphi} + J^{\sigma}$ in the obvious manner, by taking the tachyon 
and inflaton parts of $\Upsilon_1,\Upsilon_2,\Upsilon_3,\gamma$.

In appendix B we prove the identity (see eqn. \ref{gamma_sigma2})
\begin{eqnarray*}
   \gamma_\sigma &=&   
   -\frac{\kappa^2}{2}\left(\partial_i\done\sigma_A\partial^i\done\sigma^A\right)
  \nonumber \\
   &-& 3\kappa^2\Linv \partial_i\left(\Lap\done\sigma_A\partial^i\done\sigma^A\right)
\end{eqnarray*}
which is analogous to the result for a real tachyon field, derived in \cite{BC}.


We now proceed to derive the tachyon curvature perturbation.
The derivation of $\zeta^\subt_\sigma$ presented in \cite{BC} follows unmodified except,
of course, for the change in the definitions of $\Upsilon_1^\sigma$, $\Upsilon_2^\sigma$, 
$\Upsilon_3^\sigma$ and $\gamma_\sigma$.  From this point onwards we assume
that $\epsilon,|\eta| \ll 1$.  The leading contribution to the tachyon curvature
perturbation is
\begin{eqnarray*}
  \zeta_\sigma^\subt &\cong& \frac{1}{\epsilon}\int_{\tau_i}^{\tau}d\tau'
  \left[ - \frac{\Upsilon_1^\sigma}{\sH(\tau')} + 
  \frac{1}{3}\frac{\Upsilon_3^\sigma}{\sH(\tau')}\right.
  \\
  &-& \left. \frac{2}{3}\frac{\sH(\tau')^2}{\sH(\tau)^3}\Upsilon_3^{\sigma}\right]
\end{eqnarray*}
Now, using equations (\ref{Upsilon1sigma}) and (\ref{Upsilon3sigma}) we can write this in 
terms of the tachyon fluctuation $\done\sigma$ as
\begin{eqnarray}
  \zeta^\subt_{\sigma} &\cong& \frac{\kappa^2}{\epsilon}\int_{-1/a_iH}^{\tau}d\tau' \left[
                               \frac{\done\sigma'_A\done\sigma'^A}{\sH(\tau')}  \right.
 \nonumber \\ 
&-& \frac{\sH(\tau')^2}{\sH(\tau)^3}\left( 
  \done\sigma'_A\done\sigma'^A \right. \nonumber \\
&-& \left. \left. a^2 \frac{\partial^2 V}{\partial\sigma_A\partial\sigma_B}
  \done\sigma_A\done\sigma_B \right) \right]
\label{zeta_sigma}
\end{eqnarray}
The corrections to (\ref{zeta_sigma}) are either total gradients or are subleading in the
slow roll expansion.  In deriving (\ref{zeta_sigma}) we have restricted ourselves to the
preheating phase during which the fluctuations $\done\sigma_A$ grow exponentially.

Using (\ref{zeta_sigma}) the second order tachyon curavture perturbation can be computed once
the fluctuations $\done\sigma_A$ are determined.  The first order tachyon fluctuations
are described by the perturbed Klein-Gordon equation
\begin{equation}
\label{multi_comp_KG}
  \done \sigma''_A + 2\sH \done \sigma_A - \Lap\done\sigma_A 
  + a^2 \frac{\partial^2 V}{\partial\sigma_A\partial\sigma_B}\done \sigma_B =  0 
\end{equation}

\subsection{Complex Tachyon Mode Functions}

At this point we restrict our attention to the case with $M=2$ and the potential
\begin{eqnarray}
\label{complex_pot}
  V &=&\frac{\lambda}{4}\left(\sigma_A\sigma^A - v^2\right)^2 
      + \frac{g^2}{2}\varphi^2 \sigma_A\sigma^A \nonumber \\
      && + \frac{m_\varphi^2}{2}\varphi^2
\end{eqnarray}
For $\sigma^\subz_A = 0$ the mass matrix is diagonal
\begin{eqnarray*}
  \frac{\partial^2 V}{\partial \sigma_A \partial \sigma_B} 
  &=& \left( -\lambda v^2 + g^2 \varphi_0^2 \right) \delta_{AB} \\
  &\equiv& m_\sigma^2 \delta_{AB}
\end{eqnarray*}
so that the tachyon fluctuations with $A=1$ and $A=2$ evolve independently 
(see eqn. \ref{multi_comp_KG}).

As previously the quantum mechanical solutions $\done\sigma_A$ are written in terms of 
annihilation and creation operators $a_k^A,\ a_k^{A\dagger}$ in the usual
way
\beq
\label{complex_quantum}
  \done \sigma_A(x) = \int {d^{\,3}k\over (2\pi)^{3/2}}
	\,a_k^A \, \xi_k(t)\, e^{ikx}+ {\rm h.c.}
\eeq
Both components $A=1$ and $A=2$ have the same time dependence owing to the
fact that the mass matrix is diagonal.  The $\xi_k$ in (\ref{complex_quantum}) are thus
identical to the solutions of (\ref{mode}), which we have already studied.

\subsection{The End of Symmetry Breaking}

For a multi-component tachyon the condition defining $N_\star$
must be modified as $\langle \done\sigma_A\done\sigma^A \rangle(N=N_\star) = v^2 / 4$
which, for the case $M=2$, changes (\ref{end_of_inflation}) to
\[
  \left. \int \frac{d^3k}{(2\pi)^3}|\xi_k|^2 \right|_{N=N_\star} = \frac{v^2}{8}
\]

\subsection{Tachyon Curvature Perturbation}

For the potential (\ref{complex_pot}) the tachyon curvature perturbation 
$\zeta^\subt_\sigma$ decomposes into a sum of term
\[
  \zeta^\subt_\sigma = \sum_{A=1,2} \zeta^\subt_{A}
\]
where $\zeta^\subt_A$ is the contribution to $\zeta^\subt_\sigma$ coming from $\sigma_A$.
Consider, as an example, the spectrum of the tachyon curvature perturbation
\begin{eqnarray*}
  \langle\zeta^\subt_{\sigma,k_1}\zeta^\subt_{\sigma,k_2}\rangle &=& 
  \langle\zeta^\subt_{1,k_1}\zeta^\subt_{1,k_2}\rangle +
  \langle\zeta^\subt_{2,k_1}\zeta^\subt_{2,k_2}\rangle \\
  &+&   \langle\zeta^\subt_{1,k_1}\zeta^\subt_{2,k_2}\rangle +
          \langle\zeta^\subt_{1,k_2}\zeta^\subt_{2,k_1}\rangle
\end{eqnarray*}
Because the annihilation/creation operators $a^1_k$ and $a^2_k$ are independent
the cross-terms on the last line do not contribute to the connected part of the
correlation function.  This means that
\[
 \langle\zeta^\subt_{\sigma,k_1}\zeta^\subt_{\sigma,k_2}\rangle =
  2\, \langle\zeta^\subt_{1,k_1}\zeta^\subt_{1,k_2}\rangle
\]
The quantity $\langle\zeta^\subt_{1,k_1}\zeta^\subt_{1,k_2}\rangle = 
\langle\zeta^\subt_{2,k_1}\zeta^\subt_{2,k_2}\rangle$ will be identical
to the $\langle\zeta^\subt_{\sigma,k_1}\zeta^\subt_{\sigma,k_2}\rangle$
which we have already computed.
We see, then, that the effect of having a complex tachyon field 
(as opposed to a real field) is to multiply $f_L$ and $f_{NL}$
by a factor of $2$ and also to slightly reduce $N_\star$.  The net
change in $f_L$, $f_{NL}$ is order unity and the new exclusion plots
is difficult to visually distinguish from figure \ref{figN}.
This justifies our previous claims that our constraints do not change significantly
when one considers a complex tachyon field.

\section{Conclusions}
\label{VII}

In this paper we have studied the evolution of the second order curvature perturbation 
during tachyonic preheating at the end of hybrid inflation.  We have found that, depending
on the values of certain model parameters, two interesting effects are possible:
\begin{itemize}
  \item Preheating generates a scale-invariant contribution to the curvature perturbation.  In
        this case significant nongaussianity can be generated during preheating and the model
        is even constrained by producing too high a level of nongaussianity.
  \item Preheating generates a nonscale-invariant contribution to the curvature perturbation
        with spectral index $n=4$.  In this case the strongest constraint comes from the 
        distortion of the power spectrum and no significant nongaussianity can be produced.
\end{itemize}
In both cases one typically requires fairly small values of the dimensionless couplings
$g,\lambda$ in order to obtain a strong effect.  Note that a small coupling $g$ does not
require fine tuning in the technical sense, since $g^2$ is only multiplicatively 
renormalized: $\beta(g^2) \sim \mathcal{O}(g^2\lambda,g^4) / (16\pi^2)$.  That is, if $g$
is small at tree level then loop corrections do not change its effective value significantly.

We have applied our constraints on hybrid inflation to several popular models: brane inflation,
D-term inflation and F-term inflation.  In the case of brane inflation we have found that
significant nongaussianity from preheating is possible for sufficiently small values of the
warp factor.  For both D- and F-term inflation we have shown that no nongaussianity is produced
during preheating, however, we still put interesting constraints on the model due to the 
distortion of the spectrum by nonscale-invariant fluctuations.

We have also generalized the results of \cite{BC} to the case of a complex tachyon field,
confirming our previous claims that this modification does not significantly alter our
exclusion plots.

We should note that the model of hybrid inflation considered here always gives a small blue
tilt to the spectral index, $n > 1$, which is disfavoured by recent data \cite{WMAP3}.
One avenue for future study \cite{inprog} is to generalize our results to the case of 
inverted hybrid inflation \cite{inverted} which always gives $n < 1$.

\bigskip{\bf Acknowledgments.}  

This work was supported by NSERC of Canada and FQRNT of Qu\'ebec.  
We are grateful to M.\ Sakellariadou for helpful discussions and
suggestions concering our results for the complex tachyon.
We thank L.\ Leblond for alerting us to an error in the original
manuscript.  
We also wish to thank R.\ Brandenberger, C.\ Byrnes, K.\ Dasgupta,
J.\ Garcia-Bellido, A.\ Linde, J.\ Magueijo, G.\ D.\ Moore, G.\ Rigopoulos, 
T.\ Riotto, P.\ Shellard, G.\ Shiu, B.\ van Tent and F.\ Vernizzi for enlightening 
discussions and correspondence.  


\renewcommand{\theequation}{A-\arabic{equation}}
\setcounter{equation}{0}
\section*{APPENDIX A: The Matching Time $N_k$}
\label{A}

The matching time $N_k$ which determines the boundary between large- and small-scale behaviour
of the mode functions (\ref{solns}) is determined by the transcendental equation
\begin{equation}
  |N_k|e^{2 N_k} = \frac{\hat{k}^2}{c}
\end{equation}
The solutions may be written exactly in terms of the branches
of the Lambert-W functions.  In the region $\hat{k} < \sqrt{c / (2 e)}$ the 
solution is triple-valued (see figure 1 of \cite{BC}) and may be written as
\begin{equation}
\label{Nk_smallk}
    N_k = \left\{ \begin{array}{lll}
        \frac{1}{2} W_{-1}\left(-\frac{2\hat{k}^2}{c}\right) & \mbox{for the branch with 
                                                                               $N_k < -1$};\\
        \frac{1}{2} W_{0}\left(-\frac{2\hat{k}^2}{c}\right) & 
                                                         \mbox{for the branch with 
                                                                           $-1 < N_k < 0$};\\
         \frac{1}{2} W_{0}\left(+\frac{2\hat{k}^2}{c}\right) &  \mbox{for the branch with
                                                                                  $N_k > 0$}.
  \end{array} \right. 
\end{equation}
In the region $\hat{k} > \sqrt{c / (2e)}$ the solution is single valued and can be written
as
\begin{equation}
\label{Nk_largek}
  N_k = \frac{1}{2} W_0\left(+\frac{2\hat{k}^2}{c}\right)
\end{equation}

One may derive some asymptotic expressions for $N_k$ in various regions of interest.
When $|N_k| \gg 1$ we have
\begin{equation}
\label{Nk_asym_large}
  N_k \cong \ln\left(\frac{\hat{k}}{\sqrt{c}}\right)
\end{equation}
which describes $N_k$ at $\hat{k} \gg \sqrt{c / (2e)}$ and also the lower branch of $N_k$
at $\hat{k} \ll \sqrt{c / (2e)}$.  For $\hat{k} \lsim \sqrt{c / (2e)}$ there are two more
branches of the solution with approximate behaviour
\begin{equation}
\label{Nk_asym_small}
  N_k \cong \pm\,\frac{\hat{k}^2}{c}
\end{equation}

In our analysis we have used the approximation that $N_k$ is a single-valued function,
described by
\begin{eqnarray*}
  N_k^{\mathrm{s.v.}} &=& \frac{1}{2}\Theta\left(\sqrt{c/(2e)}-\hat{k}\right)
        W_{-1}\left(-\frac{2\hat{k}^2}{c}\right) \\
      &+& \frac{1}{2} \Theta\left(\hat{k}-\sqrt{c/(2e)}\right)
        W_0\left(+\frac{2\hat{k}^2}{c}\right)
\end{eqnarray*}
where $\Theta(x)$ is the Heaviside step function.  We have verified both numerically
\cite{BC} and analytically that the single-valued approximation does not significantly
alter our results.

\renewcommand{\theequation}{B-\arabic{equation}}
\setcounter{equation}{0}
\section*{APPENDIX B: An Identity Concerning $\gamma_\sigma$}
\label{B}

In this appendix we derive an identity concering the tachyon source term 
$\gamma_\sigma$ (\ref{gamma}):
\[
  \gamma_\sigma = \Upsilon_3^\sigma - 3\Linv\partial_\tau\Upsilon_2^\sigma
  - 6\sH\Linv\Upsilon_2^\sigma.
\]
Using equations (\ref{Upsilon2sigma}) and (\ref{Upsilon3sigma}) we can write this 
\begin{eqnarray*}
   \gamma_\sigma &=& \kappa^2\Linv\left[\frac{3}{2}\Lap(\done\sigma'_A\done\sigma'^A) \right. \\
   &-& \frac{1}{2}\Lap(\partial_i\done\sigma_A\partial^i\done\sigma^A)  \\
   &-& \frac{3}{2}a^2 \frac{\partial^2 V}{\partial\sigma_A\partial\sigma_B}
        \Lap(\done\sigma_A\done\sigma_B) \\
   &-& 3\partial_\tau\partial_i(\done\sigma'_A\partial^i\done\sigma^A) \\
   &-& \left. 6\sH\partial_i(\done\sigma'_A\partial^i\done\sigma^A) \right]
\end{eqnarray*}
and, after some algebra, we have
\begin{eqnarray}
   && \gamma_\sigma = \kappa^2\Linv\left[ 
   -\frac{1}{2}\Lap(\partial_i\done\sigma_A\partial^i\done\sigma^A) \right. \label{intermediate}
    \\ 
   &-& 3 \partial_i\left(\done\sigma''_A +  2\sH\done\sigma'_A + 
          a^2 \frac{\partial^2 V}{\partial\sigma_A\partial\sigma_B} \done\sigma_B\right)
   \partial^i\done\sigma^A \nonumber \\
   &-& \left. 3 \left(\done\sigma''_A  + 2\sH\done\sigma'_A 
        + a^2 \frac{\partial^2 V}{\partial\sigma_A\partial\sigma_B}\done\sigma_B\right)
   \Lap\done\sigma^A  \right] \nonumber
\end{eqnarray}
In deriving this equation we have used that fact that
\[
  \frac{\partial^2 V}{\partial\sigma_A\partial\sigma_B} 
  = \frac{\partial^2 V}{\partial\sigma_B\partial\sigma_A}
\]
which follows from the $O(M)$ symmetry of the theory.
The last two lines of (\ref{intermediate}) can be simplified using the equation of motion 
for the tachyon fluctuation (\ref{multi_comp_KG}) which gives
\begin{eqnarray}
   \gamma_\sigma 
  &=&   
   -\frac{\kappa^2}{2}\left(\partial_i\done\sigma_A\partial^i\done\sigma^A\right)
  \nonumber \\
   &-& 3\kappa^2\Linv \partial_i\left(\Lap\done\sigma_A\partial^i\done\sigma^A\right)
\label{gamma_sigma2}
\end{eqnarray}

\end{document}